\documentclass[11pt]{article}
\usepackage{moriond,epsfig}

\def\Tr{{\rm Tr}} 
\newcommand{\identity}{{\rlap{1} \hskip 1.6pt \hbox{1}}} 
\def\be{\begin{equation}}
\def\ee{\end{equation}}
\def\bea{\begin{eqnarray}}
\def\eea{\end{eqnarray}}

\begin{document}
\vspace*{4cm}
\title{Little Higgs and precision electroweak tests}

\author{Aldo Deandrea}

\address{Universit\'e de Lyon 1, Institut de Physique Nucl\'eaire,\\ 
4 rue E.~Fermi, F-69622 Villeurbanne Cedex, France}

\maketitle\abstracts{
I consider the low energy limit of Little Higgs models. The method consists 
in eliminating the heavy fields using their classical equations of motion in 
the infinite mass limit. After the elimination of the heavy degrees of 
freedom one can directly read off deviations from the precision electroweak 
data. I also examine the effects on the low energy precision experiments.}

\section{Introduction}
All models containing new physics are highly constrained by the 
electroweak precision tests. In this note I consider the 
electroweak precision data constraints on Little Higgs models by 
using a general method based on the effective Lagrangian approach:
one eliminates the heavy fields from the 
Lagrangian via their classical equations of motion in the limit of 
infinite mass, which means in practice that their mass must be 
much bigger than $m_W$. In this way one obtains an effective Lagrangian 
in terms of the Standard model fields, from which we can directly read off 
the deviations. 

\section{Little Higgs models}
It has been proposed~\cite{Arkani-Hamed:2001nc} 
to consider the Higgs fields as Nambu Goldstone Bosons (NGB) 
\cite{Dimopoulos:1981xc} of a global symmetry which is spontaneously 
broken at some higher scale $f$ by an expectation value. The Higgs 
field gets a mass through symmetry breaking at the electroweak 
scale. However since it is protected by the approximate global 
symmetry it remains light.  We shall consider in detail in the following a 
model of this type which exhibits an approximate $SU(2)$ custodial symmetry. 
The method is quite general and can be easily applied to other 
models. Similar ideas are discussed in \cite{othlh} for the littlest 
Higgs model and a class of other models. We study the electroweak  
precision constraints and give as an example of application the 
expression of the $\epsilon$'s parameterisation  
\cite{alt}. Note however that more complete parameterisations of the 
electroweak data should be used for a detailed study \cite{griwise}.
More details on the models and the methods used here can be found 
in \cite{Casalbuoni:2003ft}. Concerning unitarity limits of Little Higgs 
models see \cite{Chang:2003vs}.

\subsection{The littlest Higgs}
The model is based on a $SU(5)$ symmetry with a $[SU(2)\times U(1)]^2$ 
subgroup gauged.  This symmetry is broken down to $SO(5)$ by a vev of 
the order $f$.  
This vev also breaks the gauge symmetry to $SU(2)_W\times U(1)_Y$. This 
symmetry breaking patterns leads to 14 Goldstone bosons. Four of 
them are eaten up by the gauge bosons of the broken gauge group.  
The Goldstone boson matrix contains a Higgs doublet and a triplet under the unbroken SM gauge group.  
%
More details about this specific model and the corresponding notations 
can be found in Ref.~\cite{othlh,Han:2003wu}. 
 
The kinetic term for the scalar sigma model fields $\Sigma$ is given by 
\begin{equation} 
\mathcal{L}_\mathit{kin} = \frac{1}{2} \frac{f^2}{4} \Tr[D_\mu \Sigma D^\mu \Sigma]~, 
\end{equation} 
with the covariant derivative defined as 
\begin{equation} 
D_\mu \Sigma = \partial_\mu \Sigma - i (A_\mu \Sigma + \Sigma A_\mu^T)~. 
\end{equation} 
With $A_\mu$ we denote the gauge boson matrix: 
\begin{equation} 
A_\mu = g_1 W_\mu^{1a} Q^{a}_1 + g_2 W_\mu^{2a} Q^{a}_2 + g'_1 B_\mu^1 Y_1 + g'_2 B_\mu^2 Y_2~, 
\end{equation} 
where the $Q^a_i$ are the generators of the two $SU(2)$ groups and the 
$Y_i$ are the generators of the two $U(1)$ groups, respectively. After 
symmetry breaking the gauge boson matrix can be diagonalized by the 
following transformations: 
\begin{eqnarray}
W&=& s W_1 + c W_2 \qquad  W' = -c W_1 + s W_2 \nonumber \\ 
B&=& s' B_1 + c' B_2 \qquad B' = -c' B_1 + s' B_2~. 
\end{eqnarray} 
$s, c, s',$ and  $c'$ denote the sines and cosines of two mixing angles, 
respectively. They can be expressed with the help of the coupling 
constants: 
\begin{eqnarray} 
\nonumber 
c' &=& g'/g'_2 \hspace{0.5in} s' = g'/g'_1 \\ 
c &=& g/g_2 \hspace{0.5in} s = g/g_1~, 
\end{eqnarray} 
with the usual SM couplings $g,g'$, related to $g_1$, $g_2$, 
$g_1'$ and $g_2'$ by 
\begin{equation} 
\frac 1 {g^2}=\frac 1{g_1^2}+\frac 1{g_2^2},~~~~~~~ \frac 1 
{{g'}^2}=\frac 1{{g_1'}^2}+\frac 1{{g_2'}^2}~.\label{eq:10} 
\end{equation} 
 
The equations of motion for the heavy gauge bosons can now easily be 
obtained from the complete Lagrangian. 
We neglect, at the lowest order in the momenta,  
derivative contributions, i.e., the contributions from the kinetic 
energy vanish. Up to the order $v^2/f^2$ we obtain: 
\begin{eqnarray} 
W'^{\pm\mu} &=& \frac{cs}{2} (c^2 - s^2) \frac{v^2}{f^2}\, W^{\pm\mu} 
- \frac{4 c^3 s}{\sqrt{2}g f^2}\left( 
J^{\pm\mu} - 
(1-c_L) J^{\pm\mu}_3\right) \\ 
W'^{3\mu} &=& \frac{cs}{2} (c^2 - s^2) \frac{v^2}{f^2} (W^{3\mu} + \frac{g'}{g} \,B^\mu) - \frac{4 c^3 s}{g f^2} \left( 
J^{0\mu} -  s_L^2 \bar{t}_L \gamma^\mu t_L\right)\\ 
B'^{\mu} &=&  2c's' (c'^2 - s'^2) \frac{v^2}{f^2}(\frac{g}{g'} \, W^{3\mu} + B^\mu) \nonumber \\ && +   
 \frac{4 c' s' }{g' f^2} \left[(3 c'^2 - 2 s'^2) (J^\mu_\mathit{em} + J^{0\mu}) - \frac{5}{2} c'^2 
s_L^2 \bar{t}_L \gamma^\mu t_L  
-s_R^2 \bar{t}_R \gamma^\mu t_R \right]~, 
\end{eqnarray} 
where we have used the notation of Ref.~\cite{Han:2003wu} for the 
diagonalisation of the top sector. 

The 
input parameters in the analysis of the electroweak data are the 
Fermi constant $G_F$, the mass of the $Z$ vector boson $m_Z$ and 
the fine--structure coupling $\alpha(m_Z)$. In terms of the model 
parameters we obtain: 
\begin{equation} 
\frac{G_F}{\sqrt{2}} = 
\frac{\alpha \pi (g^2 + g'^2)}{2 
g^2 g'^2 m_Z^2} \left(1- c^2 
(c^2-s^2)\frac{v^2}{f^2} + 2 c^4 \frac{v^2}{f^2}- \frac{5}{4} 
(c'^2-s'^2)^2 \frac{v^2}{f^2}\right)\; . 
\end{equation} 
We define the Weinberg angle  as~\cite{Anichini:1994xx}: 
\begin{equation} 
\frac{G_F}{\sqrt{2}} = \frac{\alpha \pi}{2 s_\theta^2 c_\theta^2 m_Z^2}~. 
\label{weinberg} 
\end{equation} 
In terms of the model parameters the mass of the $Z$-boson is given by 
\begin{equation} 
m_Z^2 = (g^2 + g'^2) \frac{v^2}{4} \left[ 1-\frac{v^2}{f^2} 
\left( \frac{1}{6} + \frac{(c^2-s^2)^2}{4} 
+ \frac{5}{4} (c'^2-s'^2)\right) + 8 \frac{v'^2}{v^2}\right]~, 
\end{equation} 
whereas the $W$-mass is 
\begin{equation} 
m_W^2 = \frac{g^2 v^2}{4} \left[ 1- \frac{v^2}{f^2} 
    \left(\frac{1}{6} +\frac{(c^2-s^2)^2}{4}\right) 
       + 4 \frac{v'^2}{v^2}\right]~. 
\end{equation} 
The expression for 
the $Z$-mass can be used to determine the value of $v$ for a given 
ratio $v/f$.  

\begin{figure}{ 
\epsfig{file=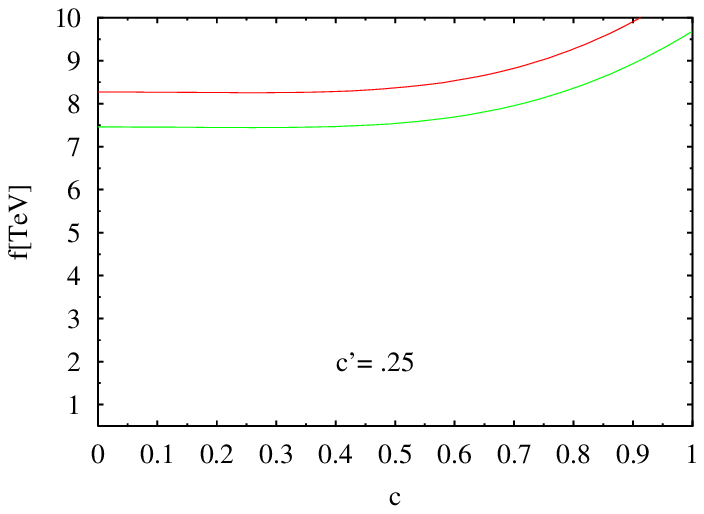,width=0.45\textwidth} 
\epsfig{file=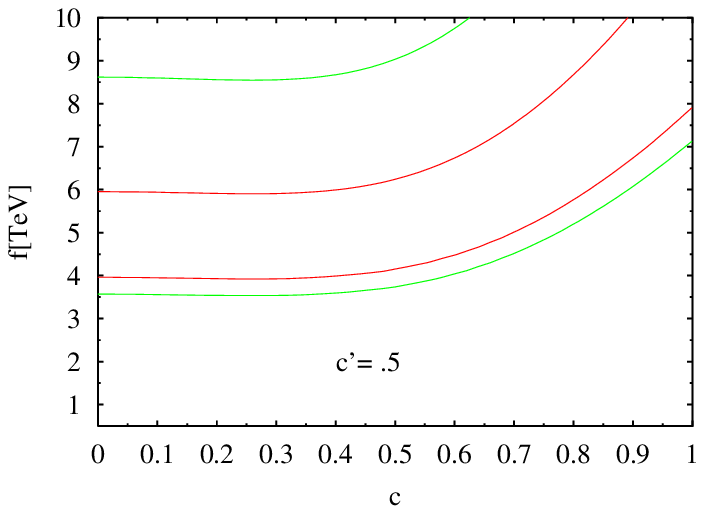,width=0.45\textwidth}
\caption{\label{epsfig}\it 90\% and 50\% CL exclusion contours in 
the plane $f$-$c$ for two values of 
the cosine of the other mixing angle $c'$ in the littlest Higgs 
model. The value of the triplet vev $v'$ is chosen using $v'/v = v/(4 f)$.
Other choices of $v'$ do not change much the above conclusions.
The allowed region lies above the bands for the left figure and inside the 
bands for the right one (90\% CL the narrowest (in red) 
and 50\% CL the largest (in green)).}} 
\label{fig:lhfcp}
\end{figure}
 
Our result for the corrections to the $\epsilon_i$ parameters to the 
order $v^2/f^2$ is given by: 
\begin{eqnarray} 
\epsilon_1 &=& - \frac{v^2}{f^2}\left( \frac{5}{4}(c'^2-s'^2)^2  
+ \frac{4}{5} (c'^2-s'^2) (3 c'^2-2 s'^2) + 2 c^4 \right) 
+ 4 \frac{v'^2}{v^2} \\ 
\epsilon_2 &=& - 2 c^4 \frac{v^2}{f^2}\\ 
\epsilon_3 &=& - \frac{v^2}{f^2}\left(\frac{1}{2 } c^2 (c^2-s^2)  
+ \frac{2}{5} (c'^2-s'^2) (3 c'^2 - 2s'^2)\frac{c_\theta^2}{s_\theta^2}\right) 
\end{eqnarray} 
Notice that the corrections, as they should, depend only on the 
parameters $c$, $c'$, $v/f$ and $v'/v$. 

The model is strongly constrained by the precision electroweak data
as can be seen in figure \ref{fig:lhfcp}.
For large values of $v/f$ the allowed 
regions are very small, whereas for small values practically the 
entire parameter space is excluded.  For large values of $v/f$ this is 
mainly due to the fact that this model exhibits no custodial symmetry 
and that it is therefore difficult to satisfy the experimental 
constraint on $\epsilon_1$ without fine tuning of the parameters. For 
small values of $v/f$ we approach the SM limit which itself is not in 
agreement with the values for the $\epsilon$-parameters. 

\subsection{Little Higgs with custodial $SU(2)$}
We now examine a ``little Higgs'' model which 
has an approximate custodial $SU(2)$ symmetry \cite{Chang:2003un}. 
The model is based on a $SO(9)/[SO(5)\times SO(4)]$ coset space, 
with $SU(2)_L\times SU(2)_R \times SU(2) \times U(1)$ subgroup of 
$SO(9)$ gauged.  
 
One starts with an orthogonal symmetric nine by nine matrix, representing a 
nonlinear sigma model field $\Sigma$ 
which transforms under an $SO(9)$ rotation 
by $\Sigma \to V\Sigma V^T$.  To break the $SO(4)$'s to the diagonal, one can 
take $\Sigma$'s vev to be 
\begin{eqnarray} 
\langle \Sigma \rangle=\left(\begin{array}{ccc} 
0 & 0 & \identity_4 \\ 
0 & 1 & 0 \\ 
\identity_4 & 0 & 0 \end{array}\right) 
\end{eqnarray} 
breaking the $SO(9)$ global symmetry down to an $SO(5)\times SO(4)$ subgroup. 
This coset space has $20 = (36-10-6)$ light scalars. Of these 
20 scalars, 6 will be eaten in the higgsing of the gauge groups down to 
$SU(2)_W \times U(1)_Y$. The remaining 14 scalars are : a single higgs 
doublet $h$, an electroweak singlet $\phi^0$, and three triplets $\phi^{ab}$ 
which transform under the $SU(2)_L \times SU(2)_R$ diagonal 
symmetry as 
\begin{eqnarray} 
h: (\mathbf{2}_L,\mathbf{2}_R) \hspace{0.5in} 
\phi^0 : (\mathbf{1}_L,\mathbf{1}_R) 
\hspace{0.5in} \phi^{ab} : (\mathbf{3}_L,\mathbf{3}_R). 
\end{eqnarray} 
These fields can be written 
\begin{eqnarray} 
\Sigma = e^{i\Pi/f}\langle\Sigma\rangle e^{i\Pi^T/f} = 
e^{2i\Pi/f}\langle\Sigma\rangle 
\end{eqnarray} 
with 
\begin{eqnarray} 
\Pi = \frac{-i}{4}\left(\begin{array}{ccc} 
0_{4\times 4} & \sqrt{2} \vec{h} & -\Phi \\ 
-\sqrt{2} \vec{h}^T & 0_{1\times 1} & \sqrt{2} \vec{h}^T \\ 
\Phi & -\sqrt{2} \vec{h} & 0_{4\times 4} 
\end{array}\right) 
\end{eqnarray} 
where the Higgs doublet $\vec{h}$ is written as an $SO(4)$ vector; 
the singlet and triplets are in the symmetric four by four 
matrix $\Phi$ 
\begin{eqnarray} 
\Phi = \phi^0 + 4\phi^{ab}\: T^{l\,a} T^{r\,b}\, , 
\end{eqnarray} 
and the would-be Goldstone bosons that are eaten in the higgsing to 
$SU(2)_W\times U(1)_Y$ are set to zero in $\Pi$.  The global 
symmetries protect the higgs doublet from one-loop quadratic divergent 
contributions to its mass.  However, the singlet and triplets are not 
protected, and are therefore heavy, in the region of the TeV 
scale. The theory contains the minimal top sector with two extra 
coloured quark doublets and their charge conjugates.  Further details 
and formulas can be found in \cite{Chang:2003un}. 
 
The kinetic energy for the pseudo-Goldstone bosons is 
\begin{eqnarray} 
{\cal L}_{kin} = \frac{f^2}{4}\Tr\left[D_\mu\Sigma D^\mu\Sigma\right] 
\end{eqnarray} 
and the covariant derivative is 
\begin{eqnarray} 
D_\mu \Sigma = \partial_\mu\Sigma +i\left[A_\mu,\Sigma \right] 
\end{eqnarray} 
where the gauge boson matrix $A_\mu$ is defined as 
\begin{eqnarray} 
A \equiv g_L W^{la}_{SO(4)} \tau^{l\,a} + g_R W^{ra}_{SO(4)} \tau^{r\,a} + g_2  W^{la} \eta^{l\,a}+ g_1 W^{r3} \eta^{r\,3}. 
\end{eqnarray} 
The $\tau^a$ and $\eta^a$ are the generators of two SO(4) 
subgroups of SO(9). For details see Ref.~\cite{Chang:2003un}. 
 
The vector bosons can be diagonalized with the following 
transformations: 
\begin{eqnarray} 
B &=&c' W^{r3} - s' W_{SO(4)}^{r3} \hspace{0.5in} 
B'= W'{}^{\,r3} = s' W^{r3} + c'  W_{SO(4)}^{r3}\\ 
W^a &=& c W^{la} +  s W_{SO(4)}^{la} \hspace{0.5in} 
W'{}^a=W'{}^{\,la} = -s W^{la} + c W_{SO(4)}^{la} 
\end{eqnarray} 
where the cosines and the sines of the mixing angles can be written in 
terms of the couplings 
\begin{eqnarray} 
\nonumber 
c' &=& g'/g_1 \hspace{0.5in} s' = g'/g_R \\ 
c &=& g/g_2 \hspace{0.5in} s = g/g_L. 
\end{eqnarray} 
Again $g$ and $g'$ are defined in terms of $g_1$, $g_R$ and $g_2$, $g_L$ respectively, as in equation (\ref{eq:10}). 

We now proceed in exactly the same way as in the previous section and 
look first at the modifications to $G_F$.  The expression for $G_F$ in 
terms of the model parameters is %
\begin{equation} 
\frac {G_F}{\sqrt{2}} = \frac{\alpha \pi (g^2 + g'^2)^2}{2 g^2 
g'^2}\left(1 + \frac{v^2}{f^2} \frac{s^2 (c^2-s^2)-s^4}{2}\right)~, 
\end{equation} 
The masses of $Z$- and $W$-bosons are given by 
\begin{equation} 
m_Z^2 = (g^2 + g'^2) \frac{v^2+4v'^2}{4} 
\end{equation} 
\begin{equation} 
m_W^2 = \frac{g^2 v^2}{4} \left(1 + 2 \frac{v'^2+{\tilde v}'^2}{v^2}\right)~. 
\label{massw} 
\end{equation} 

\begin{figure}{ 
\epsfig{file=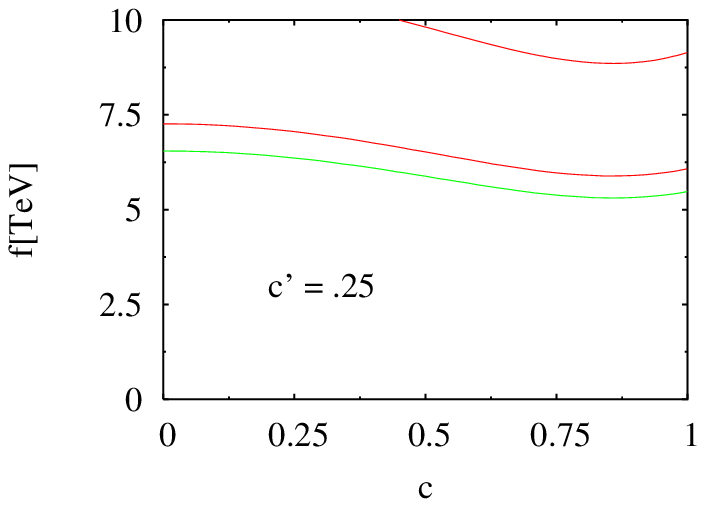,width=0.45\textwidth} 
\epsfig{file=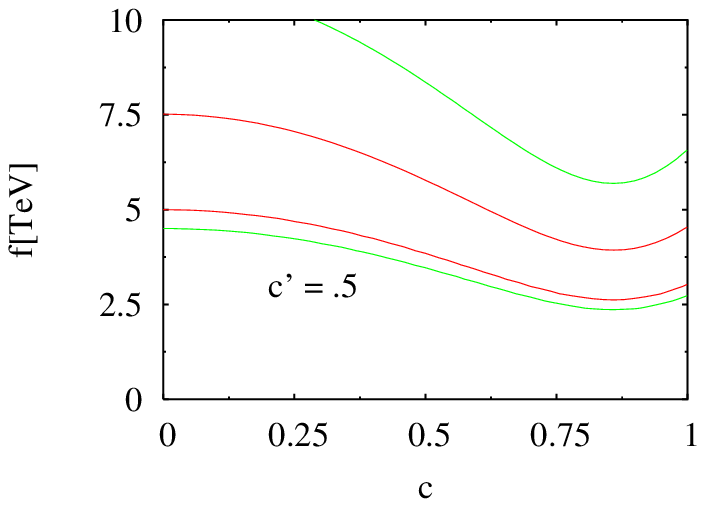,width=0.45\textwidth}
\caption{\label{epscustfig}\it Exclusion plots in 
the plane $f$-$c$ for two values of 
the cosine of the other mixing angle $c'$ in the $SO(9)/[SO(5)\times SO(4)]$ 
model. In the plot with $c'=0.25$ the value of the triplet vevs is chosen  
$v'^2/v^2 = v^2/(4 f^2)$ and ${\tilde v}'^2/v^2 = v^2/(2 f^2)$, while in the 
one with $c'=0.5$ the triplet vevs are $v'^2/v^2 = {\tilde v}'^2/v^2 = 
v^2/(4 f^2)$. The allowed region lies inside the bands (90\% CL the 
narrowest (in red) and 50\% CL the largest (in green)). }} 
\end{figure}
The $Z$ and $W$-mass only receive corrections from the triplet vevs 
($v'$ is the $Y=1$ and ${\tilde v}'$ the $Y=0$ triplet vev). This is a 
consequence of the approximate custodial symmetry of the model. 
 
The corrections to the $\epsilon$ parameters to the order $v^2/f^2$ are 
\begin{eqnarray} 
\epsilon_1 &=& \frac{v^2}{4f^2} \left[ 4 s^{\prime\, 2} 
\left( c^{\prime\, 2}-s^{\prime\, 2}\right) +2 c^2 s^2 -s^4 \right] 
+ 2 \frac{{\tilde v}'^2-v'^2}{v^2}\\ 
\epsilon_2 &=& \frac{v^2}{4 c_{2\theta}\, f^2} \Big[ 4 s^{\prime\, 
2} \left( c^{\prime\, 2}-s^{\prime\, 2}\right) c_{\theta}^2 
c_{2\theta} +2 s^2 \left( c^2 -s^2 \right)\left( c^4_\theta -3 
c^2_\theta s^2_\theta +2 c^2_\theta -s^2_\theta \right) 
\nonumber \\ 
&+&s^4 (c^4_\theta + s^4_\theta) \Big]\\ 
\epsilon_3 &=& \frac{v^2}{2 s^2_{\theta}\, f^2} 
\left[ s^2 \left( c^2 -s^2 \right) 
\left( -c_{2\theta} + 2 s_\theta^2 c_\theta^2 \right) 
-s^4 c^2_\theta s^2_\theta \right] +\frac{2 c^2_{\theta}}{v^2 s^2_{\theta}} 
\left( v'^2-{\tilde v}'^2\right)~, 
\label{epsilonscustodial} 
\end{eqnarray} 
where we have used the definition of $s_\theta$ and $c_\theta$ 
via Eq.~\ref{weinberg}. The results of the analysis are shown in 
Fig.~\ref{epscustfig} for a choiche of the model parameters. The allowed 
region lies inside the bands. As can be inferred 
from the expression of the $\epsilon$'s, a large 
value of difference of the triplet vevs spoil the custodial symmetry. Note 
however that one can argue that the two different 
triplet vevs should be similar in size and at least partially compensate 
their effects since the only violations come from the $U(1)$ coupling
which has to be smaller than the $SU(2)_R$ coupling \cite{Chang:2003un}.
Therefore the custodial violation is not a problem
for the custodial model. A precise evaluation of this effect is not possible 
in the effective theory since there are unknown order one factors in the 
radiatively generated potential.  

\section{Low energy precision data}
Precision experiments at low energy allow a 
precise determination of the $g-2$ of the muon and of the weak charge 
of cesium atoms. Concerning the $g-2$ of the muon, 
the contributions of the additional heavy particles are 
completely negligible and the dominant contributions arise 
from the corrections to the light $Z$ and $W$ couplings.
On the contrary the measure of the weak charge of cesium atoms, gives 
constraints on the little Higgs models, even if weaker that those at 
LEP energies. Parity violation in atoms is due to the electron-quark 
effective Lagrangian 
\begin{equation} 
\mathcal{L}_\mathit{eff} = \frac{G_F}{\sqrt{2}} (\bar{e} \gamma_\mu 
\gamma_5 e) (C_{1u} \bar{u} \gamma^\mu u + C_{1d} \bar{d} \gamma^\mu 
d)~. 
\label{Leff} 
\end{equation} 
The experimentally measured quantity is the so-called ``weak charge'' 
defined as 
\begin{equation} 
Q_W = -2 \left( C_{1u} (2 Z + N) + C_{1d} (Z + 2 N)\right)~, 
\end{equation} 
where Z, N are the number of protons and neutrons of the atom, 
respectively. 

\begin{figure}[ht!]{ 
\epsfig{file=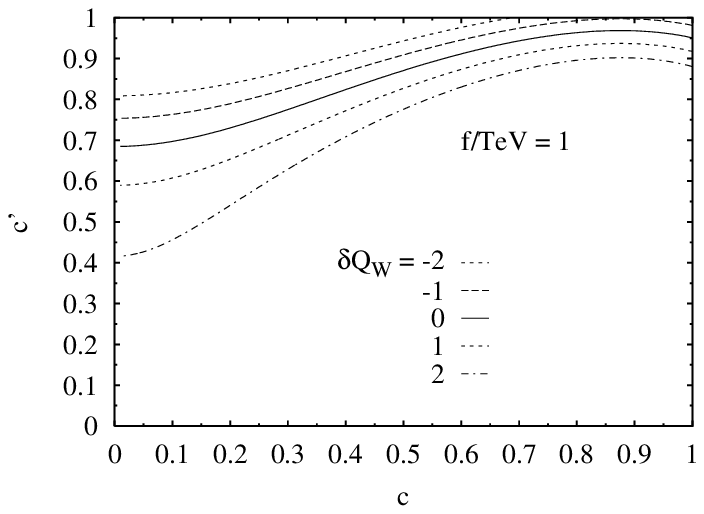,width=0.45\textwidth} 
\epsfig{file=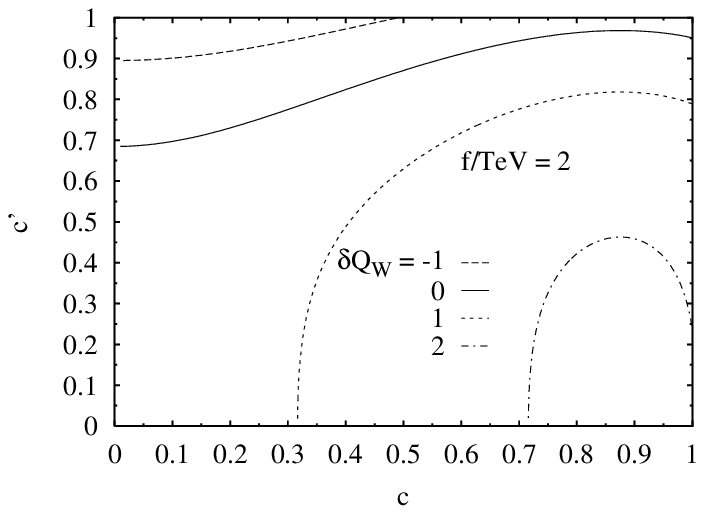,width=0.45\textwidth}\\ 
\epsfig{file=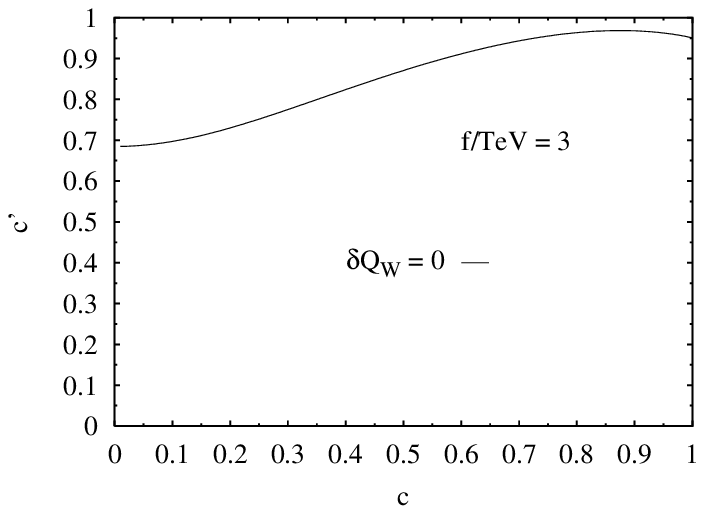,width=0.45\textwidth} 
\caption{\label{fig3cust}\it Corrections to the weak charge of cesium 
atoms as a function of $c$ and $c'$ in the 
little Higgs model with approximate custodial symmetry.}} 
\end{figure}

The effective Lagrangian, Eq.~\ref{Leff}, can be derived from the 
interaction of $Z, Z_H,$ and $A_H$ with the fermions by integrating 
out the heavy degrees of freedom.
The difference of the weak charge of Cs is shown in Fig.~\ref{fig3cust} 
for the model with approximate custodial symmetry. As experimental 
input for our analysis we have again used $m_Z, G_F,$ and $\alpha$. 
In order to discuss the weak charge result, let's consider the value  
$\delta Q_W(Cs)=1$ which is close to the present experimental central  
value. It is clear from Fig.~\ref{fig3cust} that the value  
of the high scale $f$ should be in the range of few TeV in order to obtain  
the measured deviation. The allowed scale is slightly lower in the  
custodial model with respect to the non-custodial one as the custodial model  
is closer to the standard model in its predictions. When the scale $f$ is too  
large the new physics effects become negligible.  
The scale $f$ in the few TeV range is consistent with what is expected on the  
model-building side and from the LEP data for little Higgs model.  
Obviously this result should be taken only as a first indication as the error 
on $\delta Q_W(Cs)$ is large. 
 
\section{Conclusions} 
The analysis of precision electroweak data gives rather 
stringent limits on the littlest Higgs model.  
This is mainly due to the difficulty of 
the model to accommodate for the experimental results of the $\rho$ 
parameter. In the model where custodial symmetry is approximately fulfilled, 
less fine tuning than in the littlest Higgs model is needed in order 
to satisfy the experimental constraints. Thus custodial symmetry seems to be 
an essential ingredient for realistic little Higgs models. 
Constraints from low energy precision data., i.e., $g-2$ of the muon and 
the atomic "weak charge" of the cesium, do not change the above conclusions.
For $g-2$ of the muon the 
corrections are simply too small to impose any new constraints on the 
model parameters. The actual state of precision for the weak charge 
does not allow for establishing new constraints either, even if the 
corrections are not negligible. 

\section*{Acknowledgments}
I wish to thank R. Casalbuoni and M. Oertel for the fruitful collaboration on 
which this talk is based. I also thank S. Chang, G. Marandella and A. 
Romanino for discussion.

\end{document}